%% file: main.tex
\documentclass[format=acmsmall]{acmart}
\pdfoutput=1

\renewcommand\footnotetextcopyrightpermission[1]{} 

\fancypagestyle{plain}{}
{\rfoot{plain} }

\setcopyright{none}

\usepackage{booktabs} 

\usepackage[ruled]{algorithm2e} 

\usepackage{graphicx}
\usepackage{caption}
\usepackage{subcaption}
\usepackage{enumerate}
\usepackage{amsmath}
\usepackage{booktabs}
\usepackage{multirow}
\usepackage{mathrsfs} 
\usepackage{xcolor}
\usepackage{todonotes}

\newcommand{\algname}[1] {{\fontfamily{cmtt}\selectfont {#1}}}

\begin{document}
\title{Flatter is better: Percentile Transformations for Recommender Systems}

\author{Masoud Mansoury}
\affiliation{%
  \institution{DePaul University}
  \streetaddress{School of Computing}
  \city{Chicago}
  \state{IL}
  \country{USA}}
\email{}

\author{Robin Burke}
\affiliation{%
  \institution{University of Colorado, Boulder}
  \streetaddress{Department of Information Science}
  \city{Boulder}
  \state{CO}
  \country{USA}
}
\email{robin.burke@colorado.edu}

\author{Bamshad Mobasher}
\affiliation{%
  \institution{DePaul University}
  \streetaddress{School of Computing}
  \city{Chicago}
  \state{IL}
  \country{USA}
}
\email{mobasher@cs.depaul.edu}

\begin{abstract}

It is well known that explicit user ratings in recommender systems are biased towards high ratings, and that users differ significantly in their usage of the rating scale. Implementers usually compensate for these issues through rating normalization or the inclusion of a user bias term in factorization models. However, these methods adjust only for the central tendency of users' distributions. In this work, we demonstrate that lack of \textit{flatness} in rating distributions is negatively correlated with recommendation performance. We propose a rating transformation model that compensates for skew in the rating distribution as well as its central tendency by converting ratings into percentile values as a pre-processing step before recommendation generation. This transformation flattens the rating distribution, better compensates for differences in rating distributions, and improves recommendation performance. We also show a smoothed version of this transformation designed to yield more intuitive results for users with very narrow rating distributions. A comprehensive set of experiments show improved ranking performance for these percentile transformations with state-of-the-art recommendation algorithms in four real-world data sets.

\end{abstract}

\keywords{Recommender systems, Rating distribution, Percentile transformation, Flatness}

\maketitle

\input{body-v02}

\bibliographystyle{ACM-Reference-Format}
\bibliography{sample-bibliography}

\end{document}

%% file: body-v02.tex
\section{Introduction}

Recommender systems have become essential tools in e-commerce systems, helping users to find desired items in many contexts. These systems use information from user profiles to generate personalized recommendations. User profiles are either implicitly inferred by the system through user interaction, or explicitly provided by users \cite{Adomavicius:2005,Adomavicius:2015}. In the latter case, users are asked to rate different items based on their preferences and may have individual differences in how they use explicit rating scales: some users may tend to rate higher, while some users may tend to rate lower; some users may use the full extent of the rating scale, while others might use just a small subset. \cite{Herlocker:1999a}. 

\begin{figure} 
\includegraphics[height=2in, width=3in]{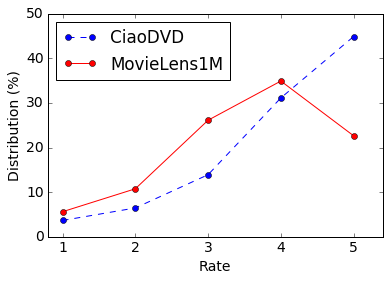}
\caption{Rating distribution of CiaoDVD and MovieLens data sets.}\label{fig1}
\end{figure}

When a user concentrates his or her ratings in only a small subset of the rating scale, this often results in ratings distributions that are skewed -- most often towards the high end of the scale. This is because items are not rated at random, but rather preferred items are more likely to be selected and therefore rated due to selection bias~\cite{marlin2007collaborative}. Figure \ref{fig1} shows the overall rating distribution of two data sets that exhibit typically right-skewed distributions. Users in the CiaoDVD data set, for example, have assigned less than 10\% of the ratings to ratings 1 and 2 and some 70\% of ratings are values 4 and 5. We can assume this is not because there are so many more good movies than bad, but rather than users are selecting movies to view that they are likely to enjoy and the ratings are concentrated among those selections. A drawback of this skew to the distribution is that we have more information about preferred items and less information about items that are not liked as well. It also means that a given rating value may be ambiguous in meaning.

As an example, assume that Alice and Bob both purchase an item \textit{X} and rate it. Alice is a user who tends to rate lower and tends to use the whole rating scale, while Bob is a user who tends to rate higher and never uses ratings at the bottom of the scale. Their profiles, sorted by rating value, are shown in Table \ref{tab:example}. After using item X, Alice is fully satisfied with it, but Bob is only partially satisfied. As a result, both rate the item X as 4 out of 5 although they have different levels of satisfaction toward that item. These ratings, while identical, do not carry the same meaning. A transformation based on percentiles, shown in the bottom rows of the table, captures this distinction well: a rating of 4 for Alice is percentile 80; whereas for Bob, the same score has a score of 50. In addition, unlike the original profiles, where the users' ratings are distributed over different ranges, these profiles span the same numerical range from 20 to 90.

\addtolength{\parskip}{-0.5mm}
\begin{table}[t]
\centering
\captionof{table}{\label{tab:example} Rating profiles with percentile transformation}
\begin{tabular}{ l | l | l}
\hline
Alice & rating & $\langle 1,1,2,2,3,3,3,4,5 \rangle$ \\
Bob & rating & $\langle 3,3,4,4,4,5,5,5,5 \rangle$ \\
\hline
Alice & percentile & $\langle 20,20,40,40,70,70,70,80,90 \rangle$ \\
Bob & percentile & $\langle 20,20,50,50,50,90,90,90,90 \rangle$\\
\hline
\end{tabular}
\end{table}

Rating normalization in neighborhood models \cite{Resnick:1994a} and inclusion of a bias term in factorization models \cite{Koren:2008a, Koren:2009a} are two common techniques for managing rating variances among users. However, these techniques adjust only for the central tendency of users' rating distributions and do not fully compensate for different patterns of rating behavior that users exhibit. On the other hand, the percentile transformation proposed in this paper takes into account the whole shape of the distribution, not only its central tendency, and therefore retains more information from the original user profile. 

Table \ref{tab:example1} shows a hypothetical rating matrix. In this table, users with different rating patterns are exhibited. Some users tend to rate lower (e.g., \textit{U1}), some users tend to rate higher (e.g., \textit{U4} and \textit{U6}), some users show normal rating pattern (e.g., \textit{U2} and \textit{U5}), and finally, some users do not show any pattern (e.g., \textit{U3}). For illustration purposes, we show how different normalization methods affect the computation of user-user similarities (in this case similarities to user $U1$). Note that for calculating the similarity values based on z-score and percentile, first we created z-score and percentile matrix from Table \ref{tab:example1}, and then we used the corresponding matrix for calculating similarity values for each technique.

\input{ratingexample.tex}

Based on users characteristics and rating patterns, certain similarity value among users is expected. For example, \textit{U1} and \textit{U4} show different behavior when providing rating to items. Based on their rating patterns, rating 3 provided by \textit{U1} can be mapped to rating 5 provided by \textit{U4}, or rating 1 provided by \textit{U1} can be mapped to rating 3 provided by \textit{U4}. Hence, a good transformation technique should be able to capture these differences in these users' behaviors. For this case, our percentile\footnote{Results are based on first index percentile transformation. The same results are observed for median and last index percentile transformation.} technique assigns high similarity value to \textit{U1} and \textit{U4}. The same result can be observed between \textit{U1} and \textit{U6}. However, original ratings and z-score technique are unable to capture these differences when calculating similarity values. In other cases, where users have normal rating patterns or do not show a specific rating pattern, our percentile technique behaves similarly to other normalization techniques. 

Moreover, above example shows that in all cases, original ratings and z-score technique behave similarly even in extreme cases when rating patterns are very different. However, our percentile technique properly takes into account those extreme cases, while behave almost similarly in normal cases.  

One can imagine the most informative rating distribution would be a flat, uniform, distribution. Users would provide ratings for items sampled uniformly across all of the items and the profiles would then represent their preferences across the whole inventory, and across all possible rating values. One way to think about the difference between the typical, skewed, distribution and a uniform one is in terms of information entropy. The worst case, a profile where every item is rated the same, carries no information that distinguishes the different items, and the assignment of a rating to an item has low entropy. A profile where the rating values are distributed across the items with equal frequency has maximum entropy.

In this paper, we formalize a rating transformation model as above that converts users' ratings into percentile values as a pre-processing step before recommendation generation. Each value associated with an item therefore reflects its rank among all of the items that the user has rated. Thus, the percentile captures an item's position within a user's profile better than the raw rating value and compensates for differences in users' overall rating behavior. Also, the percentile, by definition, will span the whole range of rating values, and as we show, gives rise to a more uniform rating distribution. To handle cases in which users use only a small part of the rating range, we also introduce a smoothed variant of the percentile transformation that preserves distinctions among users with different rating baselines.

We show that these two properties of the percentile transformation -- its ability to compensate for individual user biases and its ability to create a more uniform rating distribution -- lead to enhanced recommender system performance across different algorithms, different data sets, and different performance measures. We also show that the percentile transformation creates flatter rating distributions and that this is correlated with improved recommendation performance. 

Overall, our paper makes the following contributions:

\begin{enumerate}[\hspace{0.4cm}1.]
  \item We propose a rating transformation model that converts users' ratings into percentile values to compensate for skew in rating distributions and variances in users' rating behaviors.
  \item We empirically evaluate the proposed percentile technique using state-of-the-art recommendation algorithms on four real-world data sets. Our experiments include both overall recommendation performance and recommendation performance on long-tail items.
  \item We show the relationship between the uniformity of the rating distribution and the quality of recommendation; with flatter distributions being correlated with better recommendations.
  \item We show that the smoothed version of the transformation overcomes the issue of identical ratings in percentile and z-score transformations, and provides further improvement over the percentile alone. 
\end{enumerate}

\section{Background}

\noindent It has long been noted that users differ in their application of explicit rating scales. Resnick's algorithm, perhaps the most well-known prediction method in recommendation, normalizes ratings by user mean when computing its predictions~\cite{Resnick:1994a}. Herlocker, et al. \cite{Herlocker:1999a} used z-scores instead of absolute rating values in recommender systems and investigated its effectiveness on quality of recommendations. In this research they compared the performance of three rating normalization techniques and showed that bias-from-mean approach performs significantly better than a non-normalized rating approach and slightly better than the z-score approach in terms of mean absolute error. This result is consistent with our findings.

Kamishima in \cite{Kamishima:2010} proposed a ranking-based method that replaces the existing rating scheme with a ranking scheme. In this method, instead of rating the items, users order the items based on their preferences. Based on order statistics theory, preference orders expressed by users are converted into scores and then recommendation algorithms are applied on these scores to generate recommendations. This method proved effective, but it is not widely applicable because order-based input is rare in recommender system interfaces, and requires more effort from users than rating assignment. 

Jin, et al. \cite{Jin:2004a} compared the impact of two normalization techniques for user ratings, namely Gaussian and decoupling normalization techniques on the performance of recommender systems. This research found that decoupling normalization is more effective than Gaussian normalization. A more recent study by \cite{Kim:2016a} proposed a normalization model that learns the differences in users' rating dispositions using two phases of clustering and normalization. At the clustering phase, users are clustered based on their rating disposition and then at the normalization phase, users' ratings are normalized through predicting their rating disposition and adjusting their neighbors' ratings based on that disposition.

In the domain of trust relations in social networks, it has been shown that percentile values are more effective than absolute trust ratings. Hasan et al. in \cite{Hasan:2009a} showed that using percentile values instead of absolute trust ratings improves the accuracy of trust propagation model. They applied a method introduced by NIST\footnote{National Institute of original and Technology, http://www.itl.nist.gov/div898/handbook/prc/section2/prc262.htm} for converting predicted percentile values into trust rating in social networks.

Besides the bias in users' rating distribution, popularity bias is another well-known problem in recommender systems. Item popularity refers to the fact that depending on what recommendation algorithm we apply, inherent popularity bias in input data causes algorithms to focus their recommendations on a small set of items. Because often the items of greatest interest to users are the lesser-known ones~\cite{longtailnichesriche, park}, but these items are less common in recommendation lists -- a consequence of low quality recommendations on these items. 

Jannach et. al., \cite{Jannach2015} conducted a comprehensive set of analysis on popularity bias of several recommendation algorithms. They analyzed recommended items by different recommendation algorithms in terms of their average ratings and their popularity. While it is very dependent to the characteristics of the data sets, they found that some algorithms (e.g., \algname{SlopeOne}, KNN techniques, and ALS-variants of factorization models) focus mostly on high-rated items which biases them toward a small sets of items (low coverage). Also, they found that some algorithms (e.g., ALS-variants of factorization model) tend to recommend popular items, while some other algorithms (e.g., \algname{UserKNN} and \algname{SlopeOne}) tend to recommend less-popular items. 

Abdollahpouri et. al., \cite{himan2017} addressed popularity bias in learning-to-rank algorithms by inclusion of fairness-aware regularization term into objective function. They showed that the fairness-aware regularization term controls the recommendations being toward popular items. Also, Steck in \cite{harald2011} examined the trade-off between degrading accuracy for improving long-tail coverage. By conducting user study, they observed that adding a small bias toward long-tail items leads to better feedback from users.   


Finally, Cremonesi et. al., \cite{Cremonesi:2010a} proposed a new evaluation criterion for measuring the effectiveness of recommendation algorithms on recommending long-tail items. They compared different recommendation algorithms in terms of how accurately they recommend long-tail items to users. In fact, in their experimental setup, they measured ranking quality of recommendation outputs on long-tail items. We also follow the same evaluation criterion in the present paper to show the effectiveness of our percentile technique on long-tail items.

\section{Percentile transformation}

In statistics, given a series of measurements, percentile (or quantile) methods are used to estimate the value corresponding to a certain percentile. Given the \textit{P}\textsuperscript{th} percentile, these methods attempt to put \textit{P}\% of the data set below and (100-\textit{P})\% of the data set above. There are a number of different definitions in the literature for computing percentiles \cite{Hyndman:1996,Langford:2006}. Although they are apparently different, the answers produced by these methods are very similar and the slight differences are negligible \cite{Langford:2006}. In this paper, we use a definition from \cite{Hyndman:1996} .

The percentile value, \(p\), corresponding to a measurement, \(x\), in a series of measurements, \(M\), is computed with regard to the position of \(x\) in the ordered list \(M\), \(o(M)\), as follows: 

\begin{equation} \label{eq:percentile}
p^z(x,M)=\frac{100\times position^z(x,o(M))}{|M|+1}
\end{equation}
where \(position^z(x, o(M))\) returns the index of occurrence of \(x\) in \(o(M)\), or the position in the order where $x$ would appear if it is not present, and \(|M|\) is the number of measurements in \(M\). For more details see \cite{Hyndman:1996}.

This transformation assumes that values are distinct and there is no repetition in the series. However, with explicit rating data, we have a different situation. User profiles usually contain many repetitive ratings, and it is unclear how to specify the position of a rating. For example, in a series of ratings $v=\langle 2, 3, 3, 3, 3, 3, 5, 5, 5 \rangle$, it is not clear what the position of rating 3 should be. We could take the first occurrence, position 2, or the last occurrence 6, or something in between. 

In this work, we explore the performance of our percentile technique by taking the index of the first, median, and last occurrence of repeated ratings in the ordered vector. Hence, the parameter $z$ determines the index rule that we want to use for percentile transformation and can take values $f$, $m$, and $l$ as \textit{first}, \textit{median}, and \textit{last} index assignments, respectively. Each of these index assignments signifies a particular meaning when transforming rating profiles. The index of the last occurrence, for example, is the highest rank (most preferred) position occupied by an item with the given rating. We experiment with all index assignments and show that the rule that yields a more uniform distribution will provide greater recommendation performance\footnote{See https://github.com/masoudmansoury/percentile for the code for computing these and other transformations described in this paper.}. 

For our purposes, the entire set of ratings provided by a user \(u\) is considered a rating vector for \(u\), denoted by \(R_{u}\) with an individual rating for an item $i$, denoted as $r_{ui}$. Let $p(v,\ell)$ be the percentile mapping in Equation \ref{eq:percentile} from a rating value $v$ in a list of values $\ell$, using the first, median, and last index method. Then, the percentile value of a rating \(r\) provided by user \(u\) on an item \(i\) is computed by taking the rating $r_{ui}$ and calculating its percentile rank within the whole profile of the user. For example, based on the last index rule, for the user Bob from Table \ref{tab:example}, an item rated 3 would have percentile rank $100*2/(9+1) = 20$. We define the user-percentile function, $Per^z_u$, as follows:
\begin{equation} \label{eq:percu}
\mbox{Per}^z_u(u,i)= p^z(r_{ui}, R_u)
\end{equation}

Analogously, we can consider profiles for an item, denoted by $R_i$, to be all of the ratings provided for that item by users, and we can define a similar transformation for item profiles in which the transformation takes into consideration the rank of the rating across all ratings for that item, an item-percentile function. 
\begin{equation} \label{eq:perci}
\mbox{Per}^z_i(u,i)= p^z(r_{ui}, R_i)
\end{equation}

Note that $Per^z_u$ and $Per^z_i$ might be quite different for the same user-item pair. For example, user $x$ might be a strong outlier relative to the data set, liking an item $y$ that no one else does. $Per^z_i(x,y)$ would therefore be quite high. However, if user $x$ has a strong tendency to high ratings in general, $Per^z_u(x,y)$ might be significantly lower. This paper concentrates on the user-oriented transformation: we plan to explore the properties of the $Per^z_i$ transformation in future work.

\subsection{Measuring distribution uniformity} \label{flatness}

One of our claims in this paper is that the flatness of the rating distribution in a data set is an indicator of how well collaborative recommendation will perform, and that the percentile transformation achieves flatter distributions. In order to test this hypothesis, we need a measure of how close a rating distribution is to uniformity. 

One common technique for measuring the shape of a distribution is \textit{kurtosis}. Kurtosis is regularly used for determining the normality of a distribution. A normal distribution has a kurtosis value of 3\footnote{In some references, kurtosis is defined such that 0 reflects a normal distribution}, and a value below 3 indicates a distribution closer to uniform. Although kurtosis can be used for measuring the uniformity of a distribution, it is not a robust measure and may be misleading. Therefore, to overcome this issue, we introduce a new technique for measuring the flatness of a distribution as an alternative along with kurtosis.

To determine the flatness of a ratings distribution we calculate Kullback-Leibler divergence (KLD) between the observed rating distribution and a uniform distribution in which each rating value occurs the same number of times. If there is a discrete set of rating values $V$ (for example, {1,2,3,4,5}), then we define the flatness measure $\mathcal{F}$ as
\begin{equation}\label{eq:kldflat}
\mathcal{F}(D \parallel Q)=\sum_{v \in V} {D(v)\log\frac{D(v)}{Q(v)}}
\end{equation}
where $V$ is the set of discrete rating values in rating matrix $R$, and $D$ is the observed probability distribution over those values. $Q$ is a uniform distribution which associates a probability $1/|V|$ for each possible value in $V$ (i.e., for each $v \in V$, $Q(v)=1/|V|$). Therefore, 

\begin{equation}\label{eq:flat}
\mathcal{F}(D)=\sum_{v \in V} {D(v)\log(|V|D(v))}
\end{equation}

The $\mathcal{F}$ function measures the distance between the two distributions and hence how close the observed distribution is to the flat ideal, with a lower KLD value being indicative of a flatter distribution. 

Table \ref{tab:exflat} illustrates the flatness calculation of BookCrossing data set for original ratings. In this data set, there are ten rating values, $|V|=10$. $D(v)$ is the probability distribution over each rating values calculated as

\begin{equation}
D(v)=\frac{frequency(v)}{\sum_{v \in V}{frequency(v)}}
\end{equation}
By using equation \ref{eq:flat}, the flatness of this data set will be $\mathcal{F}=0.448$. Comparison between this flatness and the flatness of a uniform distribution (i.e., $\mathcal{F}=0$) shows that the distribution of original ratings in BookCrossing data set is far from a flat ideal. 

\input{1.KLDEX.tex}

\begin{figure}[tbh] 
\includegraphics{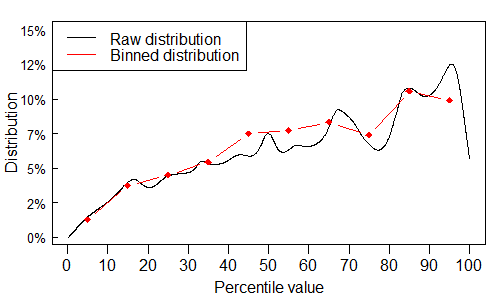}
\caption{Raw and binned percentile distributions for BookCrossing data set.\label{fig2}}
\end{figure}

The percentile and z-score transformations yield real valued ratings, unlike the original discrete ratings chosen by users in these data sets. Evaluating our flatness measure at every point in these distributions yields results that are not comparable to the original discrete distribution. 

In order to have comparable calculations of the $\mathcal{F}$ value across types of distributions, we created binned versions of the percentile and z-score distributions, using the same number of bins as present in the original ratings. In a 10-star rating system, such as found in the BookCrossing data, the rating distribution covers ten values, hence we created ten equal length bins for percentile and z-score values and aggregated each bin by its mean \footnote{As an example, we know that percentile values are between 0-100. Thus, we create ten bins each of which with the length of 10 and aggregate each bin by mean of its distribution.}. 

Figure \ref{fig2} shows the percentile distribution (last index illustrated here) and its aggregated distribution for the BookCrossing data set. The black curve is the percentile distribution and red line is its aggregated distribution with ten bins. It shows that aggregating by mean retains the shape of the percentile distribution, while being comparable to the original ratings for computing flatness.

\section{Experiments}

\noindent We evaluated the performance of percentile transformation on four publicly available data sets: BookCrossing, CiaoDVD, FilmTrust, and MovieLens. The characteristics of the data sets are summarized in Table \ref{tab:dataset}. These data sets are from various domains and have different degrees of sparsity.

\input{2.dataset.tex}

The ML1M is movie ratings data and was collected by the MovieLens \footnote{https://grouplens.org/datasets/movielens/} research group. The CiaoDVD \footnote{https://www.cse.msu.edu/~tangjili/trust.html}  includes ratings of users for movies available on DVD. The FilmTrust is a small data set collected from the  FilmTrust website \cite{guo2013}. It contains both movie ratings and explicit trust ratings.
Finally, the BX data set is a subset extracted from the BookCrossing data set \footnote{http://www.informatik.uni-freiburg.de/\textasciitilde cziegler/BX/}  such that each user has rated at least 5 books and each book is rated by at least 5 users. The ML1M has the highest density and CiaoDVD has the lowest density.

\subsection{Flatness}

To evaluate the percentile transformation for its distributional properties, we evaluated its flatness and kurtosis compared to the original ratings distribution and a distribution based on the z-score transformation over four data sets: BX, CiaoDVD, FilmTrust, ML1M.

\input{3.kld.tex}

First, we converted the original ratings into percentile and z-score values. Then, we applied the binned flatness and kurtosis measures described above to these data sets to evaluate the transformations for their distributional properties. Table \ref{tab:kld} shows the flatness ($\mathcal{F}$) and kurtosis ($\mathcal{K}$) values for each type of transformation on the four data sets (user profile transformation). As shown, the values for both measures are consistent across all three transformations and data sets. As anticipated, the proposed percentile model makes the rating values flatter than the original ratings and z-score values. Also, the original rating values show a flatter distribution than z-score values over all the data sets. 

Thus, the proposed percentile transformation approach reduces skew in the rating distribution over the original ratings and over z-score values. Given these results, we expect to see better recommendation performance when we use percentile values as input for recommender systems. We also expect that in most cases, using the original ratings will result in better recommendation performance than z-score values since they have lower $\mathcal{F}$ and $\mathcal{K}$ values.

\subsection{Methodology}

We performed a comprehensive evaluation of the effect of the percentile transformation on the ranking performance of a number of recommendation algorithms. Due to the nature of our percentile technique, we experimented only with algorithms that make use of rating magnitude. The percentile transformation rescales rating values without changing their relative ordering, so it will have no effect when applied to ranking-based algorithms (for example, ListRank~\cite{shi2010list}). Implicit feedback algorithms that use unary data, such as Bayesian Personalized Ranking~\cite{rendle2009bpr}, would also be inappropriate to use with percentile transform because they use binary interaction information and ignore rating values. 

Our experiments included user-based collaborative filtering \cite{Resnick:1994a}, item-based collaborative filtering \cite{sarwar2001}, biased matrix factorization \cite{Koren:2009a}, SVD++ \cite{Koren:2008a}, and non-negative matrix factorization \cite{lee:2001}
However, in this paper, for the purpose of presentation, we only report results on biased matrix factorization (\algname{BiasedMF}) and \algname{SVD++}\footnote{Results on all algorithms and datasets are available at https://github.com/masoudmansoury/percentile.}. Results on other algorithms in some cases were similar -- showing improved nDCG performance, although the details and significance vary -- and in some cases which our percentile technique did not improve the performance of recommendations, the results for all three input values were the same. For instance, results produced by \algname{UserKNN} on all datasets for all input values were the same, however, our percentile technique produced better ranking performance by \algname{ItemKNN} on BX and ML1M\footnote{The same result is also observed on \algname{NMF}.}.


We performed 5-fold cross validation, and in the test condition, generated recommendation lists of size 10 for each user. Then, we evaluated nDCG\footnote{We also evaluated precision, recall, and F-measure, also finding significant improvement in these metrics.} at list size 10. Results were averaged over all users and then over all folds. A paired t-test was used to evaluate the significance of results and based on paired t-test, the results shown in bold are statistically significant with a p-value of less than 0.05.

Before reporting on the results here, we performed extensive experiments with different parameter configurations for each algorithm and data set combinations. To determine sensible values for parameters, we followed the settings reported in the literature. In factorization models, for instance, we approximately set the number of factors and iterations based on the density of the data set and convergence of the loss function, and we tested these parameters for sensitivity. We performed a grid search over \textit{bias\footnote{User, item, implicit feedback, and overall bias terms.}} $\in \{0.0001, 0.001, 0.005, 0.01\}$, \textit{factor} $\in \{50, 100, 150\}$, \textit{iteration} $\in \{30, 50, 100\}$, and \textit{learning rate} $\in \{0.0001, 0.001, 0.005, 0.01\}$. 
Results of extensive experiments show that, in general, across on all settings, our percentile technique works significantly better than original ratings and z-score values.  



\subsection{Results}

\input{ndcg_table.tex}

We include results for ten experimental conditions: two recommendation algorithms evaluated over five different inputs: the original ratings, the results of the three percentile transformations, and the results of the z-score transformation. Table \ref{tab:res1} shows the results for all the data sets and both algorithms, reporting the best-performing configuration for each dataset, algorithm, and input value.\footnote{We used LibRec 2.0 for all experiments \cite{Guo2015}.}


Results in Table \ref{tab:res1} show that the percentile technique produces recommendations that are consistently better than original rating and z-score values over all the recommendation algorithms and data sets except for $\mbox{Per}^l_u$ as input for \algname{BiasedMF} on CiaoDVD data set. On the densest data set (ML1M), the average improvement by our percentile technique on \algname{BiasedMF} is 33\% and on \algname{SVD++} is 7\%. The improvement on the FilmTrust dataset is 268\% and 66\%, on the CiaoDVD dataset is 182\% and 95\%, and on BX dataset is 58\% and 48\%, respectively. Also, our results show that, in most cases, the original ratings outperform z-score transformation, which is consistent with our flatness hypothesis.

\subsection{Flatness analysis} \label{flatanalysis}

We hypothesize that a transformation that produces a flatter distribution will compensate for skew in the rating distribution and generate improved recommendation performance. As we have seen, the percentile transformation generally leads to better performance and to flatter distributions, and the less-flat z-score transformation has lower performance.

We examined this phenomena using five types of inputs for recommender system: original ratings, first, median, and last percentile values, and z-score values.\footnote{Because a limitation in LibRec, z-score values are shifted to positive values by the addition of an offset.} We examined the $\mathcal{F}$ and $\mathcal{K}$ values for the training data under the different transformations and computed correlation against the recommendation performance using nDCG@10. 

\input{correlation_table.tex}

Table \ref{tab:correlation} shows the correlation between $\mathcal{F}$ and $\mathcal{K}$ values of each input values (i.e., original rating, first index percentile values, median index percentile values, last index percentile values, and z-score values) and nDCG@10 of recommendation algorithms with those input values. It clearly shows significant negative correlation between performance and divergence from uniformity. (Note that a low $\mathcal{F}$ and $\mathcal{K}$ values correspond to a flatter distribution.) The flatter distributions (closer to zero for $\mathcal{F}$ and below 3 for $\mathcal{K}$) yield better performance for all three algorithms across all data sets. Except for the $\mathcal{F}$ value of FilmTrust on \algname{SVD++} and the $\mathcal{K}$ value of CiaoDVD on \algname{BiasedMF}, all of the observed correlations between $\mathcal{F}$ / $\mathcal{K}$ and nDCG are between -0.99 and -0.70, indicative of a strong inverse relationship: in general, flatter distributions give better algorithmic performance.


\subsection{Long-tail performance}

\input{ndcg_longtail_table.tex}

In this section, we examine the performance of recommendation algorithms on recommending long-tail items for different input values. To do this, we follow the methodology in \cite{Cremonesi:2010a} for analyzing item popularity. In this methodology, for each user in test set, a list of items will be recommended, and then ranking quality will be measured only on long-tail items in the recommended list. The main goal of this methodology is to measure the effectiveness of a recommendation algorithm in recommending long-tail items.   

For this evaluation, we need to determine the long-tail items from training data. To do this, we create cumulative popularity list of items sorted from most popular to less-popular items, then we define a cutting point such that it divides the items into short-head and long-tail items. For experiments in this paper, we used cutting point of 20\%, meaning that 20\% of most popular items are considered as short-head items and the rest of less popular items are considered as long-tail items.

Table \ref{tab:reslongtail} shows the performance of recommendation algorithms on long-tail items for different input values. As shown in this table, some version of the percentile transform significantly outperforms all other input values for each data set / algorithm combination in terms of nDCG@10. Only in three of the 24 conditions are the improvements not significant: CiaoDVD when $\mbox{Per}^l_u$ is used as input for \algname{BiasedMF} and on ML1M when $\mbox{Per}^f_u$ is used as input. 

\section{Smoothed transformation}

A drawback of the percentile transform is the handling of a \textit{uniform} user profile that consists entirely of identical ratings, for example, $\langle3,3,3,3,3,3\rangle$. When a user rates every item with the same rating values, it is hard to determine user's preferences and attitudes: if the user is generous (tends to rate highly), a rating value of 3 can be interpreted as \textit{dislike}, while if user is stingy (tends to rate low), the same value can be interpreted as \textit{like}. But in the absence of a rating distribution for a given user, it is impossible to tell which assumption is correct\footnote{Note that this issue can be even more problematic for some other transformation techniques: z-score, for example, is undefined for uniform profiles.}.

\begin{figure*}[!tbp]
  \begin{subfigure}[b]{0.3\textwidth}
    \includegraphics[width=\textwidth]{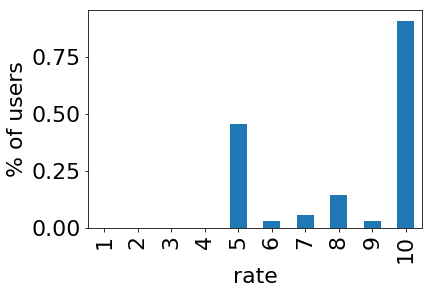}
    \caption{BX}
    \label{fig:f1}
  \end{subfigure}
  \hfill
  \begin{subfigure}[b]{0.3\textwidth}
    \includegraphics[width=\textwidth]{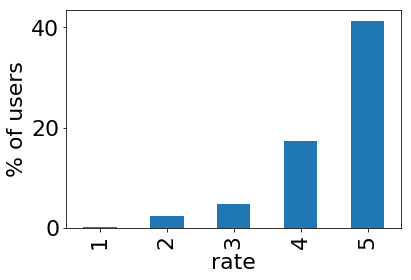}
    \caption{CiaoDVD}
    \label{fig:f2}
  \end{subfigure}
  \hfill
  \begin{subfigure}[b]{0.3\textwidth}
    \includegraphics[width=\textwidth]{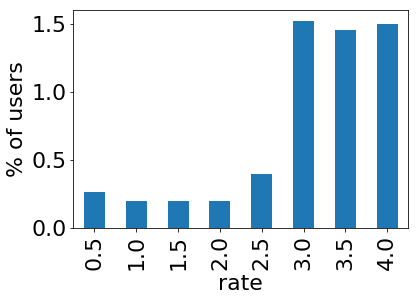}
    \caption{FilmTrust}
    \label{fig:f2}
  \end{subfigure}
  \hfill
  \caption{Percentage of users who provided identical ratings.} \label{identical-user}
\end{figure*}

Figure \ref{identical-user} shows the percentage of users with uniform profiles at different rating values in three data sets\footnote{There are no uniform user profiles in ML1M.}: BX, CiaoDVD, and FilmTrust. In CiaoDVD as the sparsest data set, more than half of the users have uniform profiles, with almost 40\% rating all items at 5. These profiles provide little information for a recommendation algorithm beyond the implicit association of user and item. 

To overcome the problem of uniform profiles, we introduce the notion of a \textit{smoothed} percentile transformation. Our inspiration for this method is additive (Laplace) smoothing as commonly found in naive Bayes classification. The effect of additive smoothing is to shrink probability estimates based on counts towards a uniform probability; here our goal is to shrink the percentile estimate towards a uniform (flat) distribution across the rating values. To create a smoothed version of the percentile, we add a small number of artificial ratings, $k$, at each rating level. In 5-star rating system, for example, possible rating values are ${1,2,3,4,5}$, so a $k=2$ smoothed transform of the profile $\langle3,3,3\rangle$ yields the smoothed profile $\langle1,1,2,2,3,3,3,3,3,4,4,5,5\rangle$.  

After the smoothed profile is created, the percentile transformation is computed and then the artificial rating values are removed, leaving behind the altered percentiles for the original rating values. Thus, the profile consisting only of 3s, as in our example above, would have middling percentile scores, being transformed to $\langle64,64,64\rangle$, using the last index method. 
If the profile had been $\langle5,5,5\rangle$ instead, the transformed version would be $\langle93,93,93\rangle$. The effect of the smoothed transform is therefore to place the user profile in the context of the full rating scale.


We formalize our smoothed version of the percentile transformation for each index assignment as follows:

\begin{equation} \label{eq:smoothedpercentilefirst}
p^f(x,M)=\frac{100\times (position^f(x,o(M)) + (k \times (index(x)-1)))}{|M|+(|R|*k)+1}
\end{equation}

\begin{equation} \label{eq:smoothedpercentileceil}
p^m(x,M)=\frac{100\times (position^m(x,o(M)) + (k \times (index(x)-1)) + k/2)}{|M|+(|R|*k)+1}
\end{equation}

\begin{equation} \label{eq:smoothedpercentilelast}
p^l(x,M)=\frac{100\times (position^l(x,o(M)) + (k \times index(x)))}{|M|+(|R|*k)+1}
\end{equation}
where $index(x)$ returns the index of rating $x$ in the full list of rating values available in the application. In rating system such as \{0.5,1,1.5,2,2.5,3\}, for example, $index(1)=2$ or $index(2.5)=5$. $|R|$ is the number of rating values available to users (i.e., in 5-star rating system, $|R|=5$).

\input{ndcg_smoothed_table.tex}

We repeated our prior experiments using these smoothed transforms, achieving the results shown in Table \ref{tab:res3}. On the FilmTrust data set, the smoothed percentile showed significantly improvement over the percentile technique particularly on \algname{BiasedMF} algorithm. On BX data set, results are only slightly better than percentile values. One might attribute this result to the fact that there are few uniform profiles ratings in BX data set. However, although ML1M does not have any users with uniform profiles, the smoothed percentile showed significant improvement over the percentile technique, indicative of effectiveness of smoothing even on non-uniform profiles.

On the CiaoDVD data set, we expected significantly better results due to high number of users who provided identical ratings. However, the improvement by smoothed percentile is only slightly better than percentile transform. One possible reason for this result is because most of the users who provided identical ratings are cold-start users with few items rated. 

\section{Conclusions}

In this paper, we presented a rating transformation model that converts rating values to percentile values as a pre-processing step before model generation. This technique addresses two well-known problems in ratings distributions in recommender systems: the problem of user rating bias, due to variation in rating behavior, and the problem of right-skew, due to the selection bias towards preferred items. This simple pre-processing step produces improved recommendation ranking performance across multiple data sets, multiple algorithms, and multiple evaluation metrics. 
In addition, we introduced the \textit{smoothed percentile} transformation to overcome the problem of identical ratings in users profiles. Experimental results showed that the smoothed percentile technique can improve recommendation performance beyond the percentile technique alone, even in cases where uniform profiles are not present. 

In introducing these transformations and demonstrating their benefits for recommender system performance, we also introduced the concept of distribution flatness and produced suggestive evidence that distributional flatness may be a good indicator of the benefits of such rating transformations: flatter, indeed, seems to be better when it comes to rating value distributions for recommendation.

In future work, we plan to conduct additional experiments with the percentile transform, particularly the item-based version of the transform, which was introduced here but for which no result were presented. Early experiments indicate that on algorithms that are item-oriented (for example, the Sparse Linear Method~\cite{Ning2011}), the item-oriented version of the transform is more appropriate. 

We also plan to explore alternative approaches to enhancing the flatness of user profiles including negative sampling. Negative sampling has been shown to improve classification accuracy when the evidence is biased~\cite{goldberg2014word2vec}. For example, rather than adding artificial ratings just for the percentile computation and removing them afterwards, it may be useful to sample items with different average rating values and use them to augment uniform user profiles. This would have the effect of smoothing such low-information profiles both towards flatness and towards the population average for item preferences.




%% file: ratingexample.tex
\begin{table}[t]
\centering
\captionof{table}{\label{tab:example1} An example of user-item matrix}
\begin{tabular}{ l | lllllllllll | rrr}

 & \multirow{2}{*}{\bf{I1}} & \multirow{2}{*}{\bf{I2}} & \multirow{2}{*}{\bf{I3}} & \multirow{2}{*}{\bf{I4}} & \multirow{2}{*}{\bf{I5}} & \multirow{2}{*}{\bf{I6}} & \multirow{2}{*}{\bf{I7}} & \multirow{2}{*}{\bf{I8}} & \multirow{2}{*}{\bf{I9}} & \multirow{2}{*}{\bf{I10}} & \multirow{2}{*}{\bf{I11}} & \multicolumn{3}{c}{Similarity to U1} \\\cline{13-15}

 &  &  &  &  &  &  &  &  &  &  &  & rating & z-score & percentile \\\cline{13-15}

\hline
\bf{U1} & 1 & 1 & 1 & - & 1 & 1 & - & 2 & 2 & 3 & 3 & - & - & - \\
\bf{U2} & 1 & 2 & 3 & - & - & - & 3 & 4 & - & 5 & 5 & 0.914 & 0.914 & 0.916 \\
\bf{U3} & - & - & - & - & 1 & 3 & - & 2 & 5 & - & 4 & 0.567 & 0.567 & 0.567 \\
\bf{U4} & 1 & 4 & 4 & - & 4 & 5 & - & 5 & 5 & 5 & 5 & 0.606 & 0.612 & 0.966 \\
\bf{U5} & 3 & 3 & - & 3 & 2 & 2 & - & 2 & 4 & 5 & - & 0.734 & 0.758 & 0.571 \\
\bf{U6} & 5 & 5 & 5 & - & 5 & 5 & - & 2 & 2 & 4 & 4 & -0.531 & -0.549 & -0.933 \\
\end{tabular}
\end{table}

%% file: 1.KLDEX.tex
%
%
%

\captionsetup[table]{skip=4pt}
\begin{table}[t!]
\small
\centering
\captionof{table}{Flatness calculation of BookCrossing data set.} \label{tab:exflat}
\begin{tabular}{lrrr}
\toprule
 rating & frequency & $D(v)$ & $\log(|V|D(v))$\\
 \midrule
 1 & 349 & 0.0029 & -3.5275 \\
 2 & 606 & 0.0051 & -2.9757 \\
 3 & 1,300 & 0.0109 & -2.2125 \\
 4 & 1,944 & 0.0164 & -1.8101 \\
 5 & 11,322 & 0.0953 & -0.0481 \\
 6 & 8,934 & 0.0752 & -0.285 \\
 7 & 19,776 & 0.1665 & 0.5096 \\
 8 & 29,233 & 0.2461 & 0.9005 \\
 9 & 21,221 & 0.1786 & 0.5802 \\
 10 & 24,113 & 0.203 & 0.7079 \\
\midrule
 & & \multicolumn{2}{c}{\large{$\mathcal{F}$=0.448}} \\
\bottomrule
\end{tabular}
\end{table}


%% file: 2.dataset.tex
\captionsetup[table]{skip=4pt}
\begin{table}[t!]
\small
\centering
\captionof{table}{Statistical properties of data sets} \label{tab:dataset}
\begin{tabular}{lrrrrr}
\toprule
 Dataset & \#users & \#items & \#ratings & density & ratings \\
 \midrule
 BX & 7,033 & 9,441 & 118,798 & 0.179\% & 1-10 \\
 CiaoDVD & 17,595 & 16,113 & 72,042 & 0.025\% & 1-5 \\
 FilmTrust & 1,508 & 2,071 & 35,497 & 1.137\% & 0.5-4.0 \\
 ML1M & 6,040 & 3,706 & 1,000,209 & 4.468\% & 1-5 \\
 
\bottomrule
\end{tabular}
\end{table}

%% file: 3.kld.tex
\captionsetup[table]{skip=4pt}
\begin{table}[t!]
\small
\centering
\captionof{table}{Flatness ($\mathcal{F}$) and kurtosis ($\mathcal{K}$) of rating distribution} \label{tab:kld}
\begin{tabular}{llrrrrr}
\toprule
 Dataset & method & rating & z-score & $\mbox{Per}^f_u$ & $\mbox{Per}^m_u$ & $\mbox{Per}^l_u$ \\ 
 \midrule
 \multirow{2}{*}{BX} & $\mathcal{F}$ & 0.449 & 0.661 & 0.110 & 0.120 & 0.101 \\
                     & $\mathcal{K}$ & 3.371 & 3.679 & 2.099 & 1.907 & 1.909 \\
\cline{1-7}
 \multirow{2}{*}{CiaoDVD} & $\mathcal{F}$ & 0.317 & 0.468 & 0.208 & 0.218 & 0.153 \\
                          & $\mathcal{K}$ & 3.619 & 3.781 & 2.384 & 2.343 & 2.044 \\
\cline{1-7}
 \multirow{2}{*}{FilmTrust} & $\mathcal{F}$ & 0.355 & 0.248 & 0.053 & 0.034 & 0.086 \\
                            & $\mathcal{K}$ & 3.206 & 3.317 & 1.938 & 1.831 & 1.850 \\
\cline{1-7}
 \multirow{2}{*}{ML1M} & $\mathcal{F}$ & 0.153 & 0.562 & 0.057 & 0.055 & 0.130 \\
                       & $\mathcal{K}$ & 2.648 & 2.920 & 2.078 & 1.829 & 1.961 \\
\bottomrule
\multicolumn{7}{l}{\textit{Flatness of a uniform distribution is $\mathcal{F}=0$.}} \\
\multicolumn{7}{l}{\textit{Kurtosis of a uniform distribution is $\mathcal{K}<3$.}} \\
\end{tabular}
\end{table}

%% file: ndcg_table.tex
\captionsetup[table]{skip=4pt}
\begin{table*}[t]
\small
\centering
\setlength{\tabcolsep}{3pt}
\captionof{table}{Performance of recommendation algorithms at nDCG@10 } \label{tab:res1}
\begin{tabular}{llrrrrr}
\toprule

 dataset & algorithm & rate & z-score & $\mbox{Per}^f_u$ & $\mbox{Per}^m_u$ & $\mbox{Per}^l_u$ \\
 \bottomrule
 
 \multirow{2}{*}{BX} & \algname{BiasedMF} & 0.009 & 0.008 & \bf{0.014} & \bf{0.014} & \bf{0.012} \\
 & \algname{SVD++} & 0.009 & 0.006 & \bf{0.011} & \bf{0.011} & \bf{0.010} \\

\cline{1-7}

 \multirow{2}{*}{CiaoDVD} & \algname{BiasedMF} & 0.019 & 0.005 & \bf{0.027} & \bf{0.024} & 0.016 \\
 & \algname{SVD++} & 0.009 & 0.011 & \bf{0.022} & \bf{0.019} & \bf{0.017} \\

 \cline{1-7}

 \multirow{2}{*}{FilmTrust} & \algname{BiasedMF} & 0.078 & 0.092 & \bf{0.317} & \bf{0.314} & \bf{0.302} \\
 & \algname{SVD++} & 0.049 & 0.072 & \bf{0.079} & \bf{0.087} & \bf{0.124} \\

\cline{1-7}

 \multirow{2}{*}{ML1M} & \algname{BiasedMF} & 0.116 & 0.116 & \bf{0.151} & \bf{0.156} & \bf{0.156} \\
  
 & \algname{SVD++} & 0.145 & 0.145 & \bf{0.153} & \bf{0.157} & \bf{0.153} \\





 \bottomrule
\end{tabular}
\end{table*}

%% file: correlation_table.tex
\captionsetup[table]{skip=4pt}
\begin{table*}[t]
\small
\centering
\setlength{\tabcolsep}{3pt}
\captionof{table}{Correlation between $\mathcal{F}$ / $\mathcal{K}$ and nDCG@10 for each algorithm} \label{tab:correlation}
\begin{tabular}{lrrrrr}
\toprule
 \multirow{2}{*}{dataset} & \multicolumn{2}{c}{$\mathcal{F}$} & & \multicolumn{2}{c}{$\mathcal{K}$} \\\cline{2-3}\cline{5-6}

 & \algname{BiasedMF} & \algname{SVD++} &  & \algname{BiasedMF} & \algname{SVD++} \\
 \bottomrule
 
 BX        & -0.95 & -0.88 &  & -0.94 & -0.85 \\
 CiaoDVD   & -0.73 & -0.72 &  & -0.62 & -0.87 \\
 FilmTrust & -0.96 & -0.51 &  & -0.99 & -0.74 \\
 ML1M      & -0.70 & -0.70 &  & -0.97 & -0.97 \\
 
 \bottomrule
\end{tabular}
\end{table*}

%% file: ndcg_longtail_table.tex
\captionsetup[table]{skip=4pt}
\begin{table*}[t]
\small
\centering
\setlength{\tabcolsep}{3pt}
\captionof{table}{Performance of recommendation algorithms on long-tail items at nDCG@10 } \label{tab:reslongtail}
\begin{tabular}{llrrrrr}
\toprule

 dataset & algorithm & rate & zscore & $\mbox{Per}^f_u$ & $\mbox{Per}^m_u$ & $\mbox{Per}^l_u$ \\
 \bottomrule

 \multirow{2}{*}{BX} & \algname{BiasedMF} & 0.0008 & 0.0009 & \bf{0.0018} & \bf{0019} & \bf{0.0034} \\
 & \algname{SVD++} & 0.0011 & 0.0009 & \bf{0.0024} & \bf{0.0024} & \bf{0.0043} \\

\cline{1-7}

 \multirow{2}{*}{CiaoDVD} & \algname{BiasedMF} & 0.019 & 0.005 & \bf{0.027} & \bf{0.024} & 0.016 \\
 & \algname{SVD++} & 0.009 & 0.011 & \bf{0.022} & \bf{0.019} & \bf{0.017} \\

 \cline{1-7}

 \multirow{2}{*}{FilmTrust} & \algname{BiasedMF} & 0.0581 & 0.0667 & \bf{0.1991} & \bf{0.2109} & \bf{0.2268} \\
 & \algname{SVD++} & 0.0279 & 0.072 & \bf{0.0603} & \bf{0.0696} & \bf{0.1006} \\

\cline{1-7}

 \multirow{2}{*}{ML1M} & \algname{BiasedMF} & 0.0364 & 0.0360 & 0.0359 & \bf{0.0379} & \bf{0.0389} \\
 & \algname{SVD++} & 0.0461 & 0.0461 & 0.0426 & \bf{0.0471} & \bf{0.0497} \\

 \bottomrule
\end{tabular}
\end{table*}

%% file: ndcg_smoothed_table.tex
\captionsetup[table]{skip=4pt}
\begin{table*}[t]
\small
\centering
\setlength{\tabcolsep}{3pt}
\captionof{table}{Performance of recommendation algorithms with smoothed percentile as input at nDCG@10 } \label{tab:res3}
\begin{tabular}{llrrr}
\toprule

 dataset & algorithms & $\mbox{SPer}^f_u$ & $\mbox{SPer}^m_u$ & $\mbox{SPer}^l_u$ \\
 \bottomrule

 \multirow{2}{*}{BX} & \algname{BiasedMF} & \bf{0.014} & \bf{0.014} & \bf{0.013} \\
 & \algname{SVD++} & \bf{0.014} & \bf{0.013} & \bf{0.012} \\

\cline{1-5}

 \multirow{2}{*}{CiaoDVD} & \algname{BiasedMF} & \bf{0.027} & \bf{0.027} & \bf{0.026} \\
 & \algname{SVD++} & \bf{0.019} & \bf{0.018} & \bf{0.016} \\

 \cline{1-5}

 \multirow{2}{*}{FilmTrust} & \algname{BiasedMF} & \bf{0.352} & \bf{0.345} & \bf{0.335} \\
 & \algname{SVD++} & \bf{0.081} & \bf{0.092} & \bf{0.102} \\

\cline{1-5}

 \multirow{2}{*}{ML1M} & \algname{BiasedMF} & \bf{0.157} & \bf{0.158} & \bf{0.156} \\
  
 & \algname{SVD++} & \bf{0.162} & \bf{0.166} & \bf{0.157} \\





 \bottomrule
\end{tabular}
\end{table*}